# Recent Advancements of Artificial Intelligence in Particle Therapy

Hao Peng[1, a)], Chao Wu[2, a)], Dan Nguyen[1], Jan Schuemann[3], Andrea Mairani[4], Yuehu Pu[2], Steve Jiang[1, b)]

[1] Medical Artificial Intelligence and Automation (MAIA) Laboratory, Department of Radiation Oncology, University of Texas Southwestern Medical Center, Dallas, TX, USA

[2] Shanghai Advanced Research Institute, Chinese Academy of Sciences, Shanghai, China

[3] Department of Radiation Oncology, Massachusetts General Hospital and Harvard Medical School, Boston, MA, USA

[4] Heidelberg Ion Beam Therapy Center, Heidelberg University Hospital, Heidelberg, Germany.

a) Co-first authors.
b) Author to whom correspondence should be addressed. Email: Steve.Jiang@UTSouthwestern.edu

*Abstract*—**We are in a golden age of progress in artificial intelligence (AI). Radiotherapy, due to its technology-intensive nature as well as direct human-machine interactions, is perfectly suited for benefitting from AI to enhance accuracy and efficiency. Over the past few years, a vast majority of AI research have already been published in the field of photon therapy, while the applications of AI specifically targeted for particle therapy remain scarcely investigated. There are two distinct differences between photon therapy and particle therapy: beam interaction physics (photons vs. charged particles) and beam delivery mode (e.g. IMRT/VMAT vs. pencil beam scanning). As a result, different strategies of AI deployment are required between these two radiotherapy modalities. In this article, we aim to present a comprehensive survey of recent literatures exclusively focusing on AI-powered particle therapy. Six major aspects are included: treatment planning, dose calculation, range and dose verification, image guidance, quality assurance and adaptive replanning. A number of perspectives as well as potential challenges and common pitfalls, are also discussed.**

## I. INTRODUCTIONS

Over the past several years, a new rising tide of AI has been changing our world in many fields, including healthcare. This new dawn of AI is largely attributable to the availability of big data, high-performance computing, and learning algorithms [1, 2]. AI has already entered the stage of applied technology and offers exciting opportunity for creating advanced tools in a wide spectrum of research areas in healthcare. Radiotherapy is a great example among them. Towards the goal of precise and personalized cancer treatment, the community has witnessed many technology advancements, including intensity-modulated radiation therapy (IMRT), volumetric modulated arc therapy (VMAT), image guided radiation therapy (IGRT), stereotactic body radiation therapy (SBRT), as well as particle therapy (proton/carbon/helium). In radiotherapy, a typical workflow comprises numerous complex tasks: imaging, segmentation, treatment planning, patient positioning/immobilization, treatment delivery, quality assurance and post-treatment follow-up [1-4]. All these tasks are quite labor-intensive and resource-intensive, lending themselves well to the integration with AI for boosting efficiency.

Compared to photon therapy, particle therapy has a unique feature, Bragg peak. Its depth-dose profile starts with a relatively flat region on the proximal end, followed by a sharp Bragg peak over the distal end [5-7]. This crucial feature allows particle therapy to achieve superior dose conformity and healthy tissue protection in comparison with photon therapy. In the meanwhile, this physical advantage also imposes a much more stringent requirement on delivery accuracy, as an example of "double-edged sword". For instance, if unexpected range and dose errors occur, proton beams will yield a dose deposition deviated from planned results, causing either "over-dose" or "under-dose" patterns. Driven by this challenge, a number of AI-related research lines have attracted extensive interest in the community, including converting dual-energy CT images or CBCT images to stopping power ratio maps, converting MR images to CT images for accurate dose calculation, converting CBCT images to CT images for adaptive re-planning, performing accurate and fast dose calculations, and online range and dose verification. The selection of works covered herein aims at presenting readers most current research and exemplifying the deployment of AI as a promising tool, with much emphasis placed on the rational, motivation and upcoming opportunities.

It is noteworthy that although many reviews about AI applications in radiotherapy have already been published [1,2,3,4,8,9,10], there is none exclusively focusing on particle therapy to the best of our knowledge. To fill in that gap, the focus of this review will be narrowed down to the meeting of AI with particle therapy. Some general topics applicable to both photon therapy and particle therapy will thus be excluded to maintain the focus. Meanwhile, we would like to point out: 1) this review is designed to be a state-of-the-art type review summarizing most recent devolvement, 2) AI in particle therapy itself is an emerging line of research and the comparison with non-AI techniques is often not easily found in many literatures, and 3) we intend to engage readers' broad interest in exploring AI as a potential tool, rather than delving into the technical details of different models. Out of these three considerations, we strive to make a good balance between the breadth and depth of this review.

## II. AI FOR TREATMENT PLANNING IN PARTICLE THERAPY

### A. Stopping power ratio mapping with dual-energy CT (DECT)

Accurate calculation of stopping power is an indispensable component in particle therapy. Dose calculation based on computed tomography (CT) images requires the conversion from Hounsfield Units (HU) to stopping power ratio (SPR). Over the past decade, a widely used approach is a piecewise linear fit based on the HU numbers in a single energy CT (SECT) measured with tissue surrogates [11-18]. However, using such an empirical fitting for HU-SPR calibration causes inaccuracy due to three challenges: many-to-one mapping, intrinsic fitting error and patient-specific tissue heterogeneities. To tackle these challenges, DECT has been recently utilized to improve the estimation of SPR by several research groups. Among these studies, the analytical method based on interaction cross-sections (e.g. photoelectric, Compton scattering) derived from DECT images in a voxel-wise manner has already been used in clinical routine.

Recently, several groups started to leverage AI tools for this task [19-22]. For example, Charyyev et al. proposed to generate synthetic DECT (sDECT) from SECT and then to derive SPR from sDECT [19], using a residual attention generative adversarial network (GAN). The corresponding SPR maps generated from sDECT are found to be comparable to the counterparts from original DECTs. In parallel, Wang et al. successfully integrated a cycle-consistent GAN to predict SPR maps directly from DECTs [20]. The datasets used in the above-mentioned two studies [19, 20] include 20 head-and-neck patients, and the leave-one-out cross validation was conducted. Despite the promising findings, a more careful evaluation is still necessary by including a larger number of patients and different DECT acquisition protocols. In 2022, Peng and his co-workers proposed the use of two convolutional neural networks (CNNs), UNet and ResNet, for deriving elemental concentration with DECT [21]. Its main purpose is to determine carbon and oxygen concentration, but the method can be easily applied for hydrogen. When no noise is present, two AI models are able to obtain <2% mean absolute errors and <4% root mean square errors in the derivation of carbon/oxygen concertation for brain images. Furthermore, different accuracy and noise immunity are indeed observed in the comparison between UNet and ResNet.

The rationale of these AI-based approaches is elaborated below. In essence, the role of AI is to map from one domain (CT images) to another domain (e.g. SPR, elemental concentration). Though conventional DECT-based approaches have been deployed for material decomposition and virtual enhancement, they face several well-known challenges such as ill-posed inverse problems, non-uniqueness and stability. AI-based models are able to avoid, or at least mitigate, these challenges. Furthermore, the AI models may be advantageous on the following two aspects: geometric prior and noise/artifact immunity [21]. Geometric prior is with regard to the difference between pixel-wise operation and image-wise operations. For example, either UNet or ResNet extracts image features through series of convolutional neural networks and down-sampling steps. As a result, not only local features but also global features can be extracted. Said differently, spatial information can be utilized as prior information to enforce geometrical correlation among voxels, such as organ type and/or boundary. On the other hand, noise/artifact immunity can also benefit from AI and machine learning, a process closely tied to geometric prior (e.g. serving as a regularization tool). Considering the physical process of CT imaging, x-rays pass through all voxels staying on a pathway to obtain projected data, and a reconstruction algorithm is used to reconstruct the image. As a result, voxels are not completely independent. Besides noise propagation in the backward projection, image artifacts such as beam hardening and scattering make such a process more complicated. Once the machine learning model was trained with high SNR images, low SNR images were tested and accurate derivation was still achieved, clearly suggesting good noise immunity and robustness of AI models [21].

*B. MR-based treatment planning*

Radiation therapy heavily relies on CT as the imaging modality for treatment planning. However, the delineation of clinical target volumes (CTVs) based on CT images suffers from poor soft tissue contrast. Magnetic Resonance Imaging (MRI) has been proposed as a complementary modality to address such limitation. To help avoid the complexity and inaccuracy of image registration between separately acquired CT and MR images, as well as to help reduce the workload, MR-only treatment planning has emerged as a new line of research. Because MRI signals are not directly linked to CT HU numbers, one central step is to generate synthetic CT (sCT) images that can be subsequently used for planning.

The existing methods of sCT generation can be broadly categorized into atlas-based [23-29], segmentation-based [30-34] and AI-based methods [35-43]. One challenge faced by the atlas-based methods is the handling of irregular anatomical structures in combination with registration errors. Segmentation-based methods are not only time-consuming but also prone to errors in manual contouring [24]. To address these limitations, several groups have explored various AI approaches, including dictionary learning [35], random forest [36] and deep learning methods [37-43]. Unlike the other two approaches, deep learning methods offer more power in feature extraction and image mapping, while requiring minimum human interaction. For instance, Spadea et al. adopted a U-Net using 15 pairs of MRI/CT head scans to predict HU values for each voxel [44]. Neppl et al. compared the performances of 2D and 3D U-Nets for head MRI images [45]. Very recently, the feasibility of a U-Net based method for pediatric patients with abdominal tumors was evaluated, based upon T1- and T2-weighted MR images [46]. The dosimetric difference between standard CT and sCT is reported to be within 2%, possibly caused by inter-scan differences (e.g. bowel filling).

For the time being, a majority of MR-to-CT conversion depend on paired MR and CT images, like U-Net CNN models mentioned above. However, possible misalignment between paired images can lead to errors in synthesizing CT images. To overcome this drawback, GAN has been introduced recently. For instance, Liu et al. used a 3D dense cycle-GAN to generate abdominal and pelvic sCTs [37, 38]. Samaneh et al. also demonstrated the feasibility of using a GAN with brain images [43]. To pinpoint the impact of heterogeneous tissues, Maspero et al. assessed the feasibility using a conditional GAN and a dataset comprising 60 pediatric brain MRI images [47]. In parallel, Erfani et al. adopted a 3D cycle-GAN with brain MRI images [48].

From the perspective of machine learning, the advantage of choosing GAN over CNN for this specific task is apparent. CNNs are trained by minimizing voxel-wise differences with respect to reference CT images that are rigidly aligned with MR images. As a result, voxel-wise misalignment between these two images sets will lead to the blurring of synthesized images. One distinct benefit of GAN is its capability to work with unpaired images, due to the bidirectional generator/discriminator workflow and the enforcement of cycle consistency [49]. Such a strength has been proved in several tasks in computer vision such as image generation, style transfer and realistic rendering. In the field of radiotherapy, a training set including MR-CT pairs may sometimes be difficult to obtain and a GAN-based model may play its role in two scenarios: 1) MR or CT images are scanned under different protocols, and 2) MR images are available but without paired CT images.

MRI-only treatment planning in particle therapy emerges as an active field offering several potential benefits, such as eliminating MR-CT co-registration errors, reducing CT radiation exposure, and simplifying clinical workflow. The rational for AI deployment is not that much different from section II.A, and the same limitation related to the small size of datasets also applies. Future work should consider using larger and more heterogeneous datasets for more rigorous validation. Special attention should also be paid to examine to what extent an AI-based approach outperforms conventional approaches, in terms of quantitative dosimetric metrics.

### III. AI FOR DOSE CALCULATION IN PARTICLE THERAPY

Dose calculation is a critical component in radiotherapy. With the current pace of technological advances, AI-powered dose calculation has already matured to a point of clinical translation, and will continue to receive wider attention. Differing from photon therapy, two factors make the task of dose calculation more demanding in particle therapy: the shape of Bragg peak and the physics of charged particle interaction with tissues. Two approaches for dose calculation in proton therapy have been widely studied: pencil beam algorithm (PBA) and Monte-Carlo simulation (MC) [50-52]. An important trade-off is between accuracy and efficiency. In a PBA approach, a number of pencil beams are ray-traced individually as they deposit energies inside a patient. Its advantages include easy to use, commercial availability and high-speed. In contrast, a MC approach models the step-by-step interaction of individual particles down to the very fundamental level of physics, and is believed to deliver the "gold standard" with utmost accuracy. However, this comes at the cost of heavier computational workload and longer calculation time. The limited accuracy of PBA stems from two reasons. The first reason is that it does not fully embody tissue heterogeneity. Patients are modeled as a stack of semi-infinite layers and the materials encountered by each pencil beam are assumed to be homogenous laterally. Such an assumption is not valid in highly heterogeneous sites such as head and neck, and lung. The second aspect is the impacts resulting from Coulomb scattering (with orbital electrons) and nuclear interactions, both of which are dependent on beam energy and depth. Numerous literatures have already investigated the dosimetric discrepancy between PBA and MC models [53-56]. Over the years, the utilization of graphical process units (GPUs) has also been extensively investigated in order to speed up MC-based dose calculation [57-60].

In a series of pioneering studies, the UT Southwestern group proposed the use of AI (a densely connected U-Net) for dose calculation in radiotherapy [61-63], starting with photon therapy and then expanding to proton therapy. As mentioned above, one dilemma in dose calculation is that the fast algorithms are generally less accurate, while the accurate dose engines are often time consuming. The group proposed to resolve this dilemma by exploring deep learning for the first time. In simple terms, the AI-model is expected to convert the less accurate results from a fast algorithm to the MC results with good accuracy. In one study [63], the model uses the PB doses and CT images as inputs to generate the MC doses as outputs. For a dataset of 290 patients (90 head and neck, 93 liver, 75 prostate and 32 lung), the average gamma passing rate (1 mm/1%) between the AI-predicated dose maps and the counterparts directly from MC is 92.8% (head and neck), 92.7% (liver), 89.7% (lung) and 99.6% (prostate), respectively. The calculation for a single field takes less than 4 seconds.

Along the same path, another group used a 3D CNN for head and neck cases [64]. Beam parameters such as spot weighs, spot position and beam energies were used as inputs, and the dose maps obtained through MC were used as outputs. Moreover, a transfer learning technique was used in order to speed up training and improve the generalization capability of the CNN model, though the rational of taking such step is not fully convincing in our view. Furthermore, as the authors pointed out, there exist two obstacles to be further addressed [64]: 1) multiple intermediate stages are required for pre- and post-processing, and 2) the voxel resolution was set to 4 mm due to limited GPU memory capacity, larger than a typical value in clinical settings (2-3 mm).

Recently, Neishabouri et al. proposed a Long-Short Term Memory (LSTM) recurrent neural network (RNN) model for pencil beams [65]. Relative stopping power values of tissues were used as inputs, and dose maps obtained through MC simulation were used as outputs. The authors tested the model with both digital phantoms and lung images, which exhibit a high degree of tissue inhomogeneities (e.g., interfaces between lung tissues and high-density rib cages). When compared to MC results (the benchmark), good agreement is found and the gamma-index pass rate stays between 94% and 98% for three beam energies (67.85, 104.25, and 134.68 MeV). The calculation time is about 1.5 ms for a single beamlet with a consumer GPU. How to extend the current work to a complete treatment plan comprising multiple pencil beamlets is the remaining task. Another study developed a hierarchically densely connected U-Net model [66], with dose maps calculated with a PBA and patient CT images as inputs. The outputs were the dose maps from MC simulations at different organ sites: head and neck, liver, prostate and lung, similar to [63]. The model clearly showcases its strengths over PBA with regard to multiple evaluation criteria, including Gamma index, mean square error, dose volume histogram (DVH) and dose difference. Moreover, the authors claimed that with additional efforts to improve the model efficiency (i.e., model compression), the calculation time can be further shortened.

Besides CNN and U-Net, one promising study investigated dose calculation using a discovery cross-domain GAN (DiscoGAN) [67]. The training data was generated using MC simulations. In essence, the DiscoGAN was designed to perform the mapping between two domains: beam parameters and dose, while HU values from CT images and an analytical derived stopping power (SP) kernel are incorporated as auxiliary features. The information such as beam energy and cross-section can be implicitly embodied through the SP kernel. At the heart of this process, is the capability of the AI model to capture the complicated relationship between dose and HU/SP. For a single beamlet, the calculation time is about 0.5 sec. The accuracy was quantitatively evaluated in terms of mean relative error (MRE) and mean absolute error (MAE). The mean MRE is consistently below 3% for all three sites (head and neck, thoracic, abdomen). Furthermore, no systematic deviation, either over-dose or under-dose, is found between the AI and MC approaches.

When coming to the design of AI models, the study in [67] provides a good example on two aspects: 1) choosing a suitable network model tailored for a specific task in particle therapy (e.g. the nature of generative capability), and 2) combining AI with physics to interpret the findings and potential benefits (e.g. analytically derived stopping power as prior information). The second aspect one is often overlooked in AI-related papers in our field. As the authors stated, a heuristic view may help illuminate how the AI model essentially functions. First, the SP modeling behaves quite like a PBA to yield a preliminary dose profile close to the "true" dose profile, as would be obtained by MC simulation. Second, the AI model fine-tunes the preliminary dose profile to address minor discrepancies resulting from tissue heterogeneity and physical processes (Coulomb scattering and nuclear interactions), bringing the result further closer to the "true" dose profile.

## IV. AI FOR RANGE AND DOSE VERIFICATION IN PARTICLE THERAPY

Range/dose verification in particle therapy, is a task distinct from other sections on several aspects. First, there are secondary signals induced by particle beams, resulting from unique physical processes not seen with high energy photon beams. Second, the task is closely tied to the instrumentation such as hardware design and signal processing, making it more challenging compared to a task involving software only. Third, while the adoption of AI in this research area is of great promise, it is still in the very early phase and has limited literatures available for discussion.

A major challenge in particle therapy is how to accurately monitor the location of Bragg peak and "real" dose distribution. Possible uncertainty may result from a number of factors such as beam profile, stopping power conversion (e.g. from CT images), patient positioning and anatomical changes. The underlying principle is that the spatial distribution of secondary signals correlates with dose distribution. Several types of secondary signals have been examined, including positron emitter, prompt gamma, secondary electron bremsstrahlung x-ray, acoustic wave and water luminesce. Each category has its own strengths and limitations. Among them, the use of Positron Emission Tomography (PET) has been extensively studied for detecting positron emitters (e.g. $^{11}$C, $^{15}$O), either with or without AI [68-78]. In our opinion, PET-based verification is the most promising tool to find clinical translation. From the perspective of AI and machine learning, the rational of AI deployment is highly similar among different verification approaches. To maintain the focus of this review, PET-related studies are placed with more weights and we encourage readers to delve into other applications that might be of interest to them.

### A. Positron emitter

In a series of studies, Peng and his co-workers proposed the use of multiple AI models for range/dose verification in proton therapy for the first time [68-71]. The logic behind these studies is clear: as the models get more complicated and more physics-related information are incorporated, AI unleashes its increasing power. For instance, Li et al. started with a simple forward neural network and a LSTM RNN to predict 1D dose distribution for mono-energetic beams [68]. Later, Liu et al. compared the performance of five RNN models: LSTM, bi-directional LSTM, GRU, bi-directional GRU and Seq2seq. The impact of including anatomical information (HU numbers) was also thoroughly examined. The results suggest that the bi-directional GRU structure achieves the most accurate prediction and the best generalization capability, especially with the presence of HU as features. Built upon these two studies, the team gradually enhanced the AI framework on the following four aspects [70, 71]: 1) adding stopping power along with HU as prior information to enhance noise immunity and generalization capability, 2) realizing 3D verification for both center and off-center voxels, 3) assessing spread-out Bragg peak (SOBP) cases in addition to mono-energetic cases, and 4) testing the performance of two input scenarios (reconstructed PET signals and raw positron emitter signals).

Nevertheless, one limitation of these studies is that the dataset for model training/testing was based on a single CT image. The performance for different organ sites needs to be rigorously checked, even though the framework is devised for patient-specific verification.

When conducting AI-related studies in the field of particle therapy, or radiotherapy in general, two general questions should be answered. First, what is the rational or motivation of introducing AI for a given task? Second, whether there are proven benefits in comparison to non-AI approaches? To certain degree, the answers to these two questions carry more weights than the development of an AI model itself. The studies in [68-71] exemplify how to effectively do so. For instance, the authors fully explained why the mapping from a dose profile to an activity profile (e.g. positron emitter signals) is not a trivial task. The reason is because dose profiles depend on stopping power and the medium, while the yield of proton-induced positron emitters depends on another parameter, the cross section of nuclear reaction besides stopping power and the medium. As a result, a typical activity versus depth profile exhibits complicated fluctuation patterns, distinct from a smooth dose versus depth profile. The selection of RNN-based models is due to its strength in extracting the underlying correlation between an input sequence (activity profile) and an output sequence (dose profile), while requiring a reduced number of parameters. Furthermore, the authors thoroughly explained the necessity of including HU and/or SP as extra features. When no anatomical information is incorporated, an RNN model focuses largely on global features. The inclusion of HU and SP helps the RNN model to capture local correlation. With regard to the individual contribution of HU and SP, the former one is associated with the cross-section for x-ray (i.e. information of carbon and oxygen concentration in tissues) and correlates with the activity profile in an indirect way, while SP is directly linked to the dose profile.

With regard to potential benefits relative to non-AI models, the authors provided plausible explanations as well [69-71]. For instance, it is easy to use and has the potential to provide an end-to-end solution. It extracts features automatically, different from pre-selected functions and fitting routine in kernel-based models (e.g. the convolution of a Gaussian function and a power-law function) [72-75]. It demonstrates the strength in robustness, generalization capability and noise immunity. Some quantitative comparisons are also presented in does verification [76] and range verification [72, 73, 76-78], between AI and non-AI studies. Though a direct comparison is sometimes difficult to be made, indirect comparisons like these are still conducive to other colleagues working on the same topic.

### B. Prompt gamma

During the interaction of charged particles with tissues, characteristic photons (also known as prompt gamma, PG) are emitted. Similar to PET-based verification, PG signals are highly correlated with the dose distribution [79-82]. Gueth et al. proposed an approach to detect possible discrepancies between planned and delivered dose [80], based upon a combined classifier using distal falloff and registered correlation as features. Liu et al. applied a U-Net model to tackle the same task with brain phantoms [81], selecting MC-simulated PG signals as inputs and dose maps as outputs. Recently, Schumann et al. proposed to combine a filtering procedure based on Gaussian-power law convolution with an evolutionary algorithm [82]. For all these studies, however, a number of physical factors that will degrade the quality of PG signals (i.e. counting statistics, limited spatial resolution, image artifacts), have yet to be examined.

C. *Secondary electron bremsstrahlung x-rays*

Secondary electron bremsstrahlung (SEB) x-rays can also be utilized for range and dose verification [83-87]. The prototype system has been developed and tested in both proton therapy [84] and carbon therapy [85-86]. The essential goal is how to convert SEB x-ray images to dose images, facing the presence of two challenges: limited spatial resolution and poor counting statistics. To address these challenges, Yamaguchi et al. proposed the use of two U-Net models, one for x-ray to dose conversion and the other for resolution enhancement [87]. Instead of using MC simulations, a pre-defined analytical function with parameters extracted from experimental measurements was adopted to ease the workload of data generation. Despite the promising results, it can be foreseen that a formidable obstacle exists when the AI approach is applied to heterogeneous tissues.

D. *Acoustic signals*

Numerous studies exploited thermoacoustic signals for range/dose verification [88-95]. Due to the characteristics of signal production, two types of pressure waves are emitted by the pre-peak dose deposition and the peak dose deposition. The time-of-flight (TOF) method has been investigated for a uniform water medium [88, 89]. In parallel, the time-reversal (TR) reconstruction method in both 2D and 3D heterogeneous tissues has been proposed to tackle two challenges associated with the TOF method [92, 93]: 1) complicated extraction of arrival time, and 2) incapability to provide complete dose information. However, compared with the TOF method, the TR method is much more time consuming, taking up to several minutes for 3D calculation even with GPU acceleration. To address the time constraint, Yao et al. developed an AI model (Bi-LSTM RNN) for both 2D and 3D scenarios [94, 95], which is found to be able to identify the correlation between acoustic waveforms and dose profiles, in combination with advanced signal processing techniques (e.g. Hilbert transform, wavelet decomposition, etc.). Compared to the TR approach, the AI model requires much shorter computational time (several seconds) and much less sensors for detecting acoustic waves.

The advantage of an RNN model for the task lies in its unique strength in extracting sequential correlation, as well as requiring a small number of hyper-parameters. As illustrated in [94, 95], every single dose profile actually bears a close correlation with a unique 2D map comprising multi-channel acoustic waveforms. The RNN model analyzes the characteristics of time-series signals for feature extraction. For example, one feature of particular importance is the relative time offset of amplitude peaks among individual detector channels. The authors also mentioned that to treat multi-channel signals all at once using a CNN is also a viable solution, but two potential concerns make such a choice less appealing. First, a CNN requires a much larger number of hyper-parameters to be trained (i.e. demanding more data samples for training). Second, a CNN may not provide the same strength in identifying temporal correlation as an RNN model. In spite of these potential advantages, data generation for model training remains a formidable challenge. To be specific, the propagation of acoustic waves and their shapes strongly rely on the set of physical parameters (e.g. sound speed in heterogeneous tissue). Both input and output for machine learning were so far originated from simulation, for both non-AI and AI approaches [92-95]. How closely the simulated dataset represents the signals in practice, needs to be carefully examined once a prototype system is built.

### E. Luminescence

The luminescence signal from water during particle therapy is another line of research [96-98]. The luminescence of water is produced through a similar process to Cerenkov light, which can be detected with a cooled charge-coupled device (CCD) camera. Yabe et al. used a U-Net to predict 2D dose distributions from the measured luminescence images of a water medium for both proton and carbon beams [98]. Similar to the secondary electron bremsstrahlung (SEB) x-rays described above, this approach will encounter obstacles when applied to heterogeneous and/or deep-seated tissues. Furthermore, we would like to reiterate our belief that for this type of research, as well as many other mentioned in this section, the dominating challenge is not tied to AI itself. Even a very basic AI model may suffice the requirements. Instead, the major challenge stays within other aspects such as instrumentation and signal detection. For this reason, a justification of choosing AI over a non-AI approach, as well as the plan for performance comparison between two avenues, should be offered.

### V. AI FOR CBCT IMAGE GUIDANCE IN PARTICLE THERAPY

Image guidance is another important component for precision radiotherapy, particularly towards the realization of adaptive radiotherapy. The need for adaptation in particle therapy is more critical than in photon therapy, due to its unique Bragg peak and steep dose gradient [99-101]. Any patient-related deviation from the initial treatment plan (positioning, anatomy, etc.), will cause dose inconsistences and consequently comprise treatment efficacy. Cone-beam CT (CBCT) is a widely used tool for image guidance, one already integrated with a number of proton therapy systems. CBCT has manifested its role in the verification of patient positioning, monitoring anatomical changes (intra-fractional changes), as well as adjusting a treatment plan accordingly.

However, the image quality of CBCT is not sufficient for yielding accurate HU numbers [102-106]. Over the past decade, numerous non-AI approaches have been proposed to tackle this issue, including look-up tables [102], histogram matching [103], deformable image registration (DIR) [104, 105] and improved scatter correction [106]. Recently, how to leverage the strength of AI has attracted increasing attention. The discussion in this section focuses exclusively on two aspects: CBCT to CT conversion [107-113], and CBCT to SPR mapping [114-116]. As a matter of fact, there is a very fine line between these two aspects since SPR derivation is the ultimate goal, and several literatures comprise both for completeness. Furthermore, we would like to clarify two points. First, the tasks of image conversion/synthesis and SPR derivation share many common features as the examples already presented in section II, so the mere emphasis below is placed on CBCT (including scatter correction). Second, besides image guidance, how to make a treatment adaptive needs to take into account beam-related uncertainties (spot position, monitor unit, beam energy, etc.) and replanning algorithms, a topic to be discussed in section VI.

### A. CBCT to CT image conversion

Acquiring paired CT and CBCT images is a difficult task in practice. As discussed in section II.B, the GAN is able to alleviate the need for paired CT and CBCT images. For instance, Liang et al. adopted a cycle-GAN to generate sCTs from CBCT images for head-and-neck patients [107]. Similarly, Kurz et al. evaluated this feasibility for both VMAT and proton therapy [111]. The authors conclude that the accuracy of dose calculation based upon sCTs is sufficient for VMAT, but not for proton therapy.

AI approaches have also been reported when paired CT and CBCT images are available. Hansen et al. used a 2D U-Net to correct CBCT images for both IMPT and VMAT plans [108]. For pelvic patients, the results exhibit satisfactory accuracy only for VMAT plans. In another study [109], Landry et al. trained a U-Net with three types of datasets for prostate images: (1) raw CBCT and CT images, (2) raw CBCTs and DIR-synthetic CTs; (3) raw CBCT and scatter-corrected CBCT. Such design intends to decouple the impact of scatter correction and deformation, and the third dataset achieves the best performance. Thummerer et al. used a dataset of 27 head and neck patients, containing planning CT, repeated CTs, CBCTs and MRs to train a U-Net for sCT generation [110]. The results suggest that CBCT-based sCTs have a higher degree of similarity relative to planning CTs than MR-based sCTs, and both are equally suited for daily adaptive proton therapy.

### B. CBCT for SPR mapping

The goal of this task to is to improve SPR estimation and dose calculation, similar to section II.A. Harms et al. recently proposed a cycle-GAN to extend its role from CBCT to CT conversion to CBCT to SPR mapping [114]. The authors conclude that the AI approach achieves comparable, if not superior, performance to that of a DIR method for head-and-neck patients, in terms of mean absolute error (MAE), mean error (ME), peak signal-to-noise ratio (PSNR), and structural similarity (SSIM). The MAE between CT-based and CBCT-based SPRs is $0.06 \pm 0.01$ and the ME is $-0.01 \pm 0.01$. There are two limitations in the current study [114]: 1) the empirically selected HU-SPR curve was used for data generation, which had an intrinsic inaccuracy up to a root mean squared error of 5.5%, and 2) the rational of the direct inclusion of SPR for learning demands a clearer justification.

One crucial task to improve CBCT image quality is scatter correction. Conventional scatter correction strategies require either complicated analytical models with ad-hoc assumptions, or heavy computational burdens such as the well-known scatter kernel superposition algorithm. The role of AI models for more effective scatter correction and accurate SPR derivation, is exemplified in the following two examples. Lalonde et al. evaluated the performance of a U-Net to evaluate scatter correction for 48 head and neck CBCT images [115]. Dosimetric performance and proton range were compared among three scenarios: scatter-free (ground truth obtained through MC simulation), uncorrected and scatter-corrected CBCT images. For AI-powered scatter correction, the mean HU difference with regard to the ground truth decreases from -28.6 HU (uncorrected images) down to -0.8 HU (corrected images). The root-mean square error of proton range between the ground truth and the scatter-corrected scenario is 0.73 mm. The correction for the complete image volume can be completed in less than 5 seconds. In parallel, Numora et al. evaluated a U-Net for two sites, head/neck and lung [116]. The method was compared to the conventional method, fast adaptive scatter kernel superposition (fASKS). Four figures-of-merit were selected for quantitative comparison, similar to those used in [114]. The authors surmise that the AI model not only outperforms the fASKS method with regard to all four metrics, but also is computationally more efficient. The correction for 360 projections takes less than 5 second on a PC (4.20 GHz Intel Core-i7 CPU) with a NVIDIA GTX 1070 GPU.

It is apparent that the potential benefits of AI can be harvested in both CBCT to CT conversion and CBCT-based dose calculation, particularly in the context of adaptive therapy. Future studies should focus more on the generalization and robustness of these AI approaches, offering full validations with multiple organ sites and large datasets. The fidelity of a training dataset generated through MC simulation, as well as whether the MC simulation differs from an acquisition protocol in practice, should be put under scrutiny.

## VI. AI FOR OTHER APPLICATIONS IN PARTICLE THERAPY

Here we review three niche areas in particle therapy that may also leverage the power of AI. For the time being, they are scarcely studied relative to the topics covered in other sections. Along with the advancement of AI tools and the push for precision radiotherapy, they may see ample opportunities for new AI development. When a task involves no images, deep learning models such as CNNs will not be necessary and other tools such logistics regression, support vector machine and decision trees are likely to suffice the need.

## A. Patient-specific quality assurance (QA)

QA is a critical component in radiotherapy and its main goal is to guarantee the accuracy of dose delivery [117]. Since it is not only a labor-intensive process but also prone to measurement uncertainties, AI tools are of great potential to fit in. Back in 2011, Zhu et al developed a support vector regression (SVR) model to establish the correlation between DVHs of OARs (bladder and rectum) and anatomical information (e.g. organ volumes, distance-to-target histogram) [118]. Recently, AI techniques have also been adopted as a secondary check tool for the prediction of monitor units (MUs) and dose output.

For example, before the wide spread of the pencil beam scanning mode in particle therapy, treatment planning systems did not have built-in modules to calculate MUs in the passive scatter mode. Phantom measurements are thus required to determine the field-specific dose output, a routine subjected to measurement errors and limited machine time. To address this problem, Sun et al proposed an AI-based approach to predict dose output (cGy/MU), using gantry angle and field size as features [119]. The authors conclude that all three models (Random-forest, XGBoost, and Cubist) outperform an empirical model, for a dataset comprising 1,754 treatment fields and phantom measurements. A similar work was also reported for the pencil beam scanning mode [120], in which Grewal et al. used Gaussian process regression (GPR) and shallow neural networks to predict multiple QA parameters (range, modulation, field size and output factor). Again, for a training dataset of 4,231 patient-specific QA measurements, both models outperform an empirical model.

Very recently, the use of AI for the prediction of treatment delivery errors emerges. Maes et al developed three models (linear regression, random-forest and neural network) where the planned spot parameters (e.g. spot position, monitor units and energy) were extracted from TPS as inputs, and the delivered spot parameters were extracted from log-files as outputs [121]. The dataset contained treatment plans of 20 prostate patients. An intriguing comparison presented is that in terms of standard deviation, the uncertainty in X/Y positions (difference between plan and delivery) reduces from 0.39 mm/0.44 mm, down to 0.22 mm/0.11 mm (difference between prediction and delivery). The random-forest model offers the best predictive power. Two comments deserve a mention here. First, although the preliminary feasibility of AI-based identification of delivery errors was proved, a much larger dataset is needed. Second, a logical question rises immediately whether and how such predicted error be taken into account before beam delivery.

## B. Decision support system

The idea of leveraging AI in the development of decision support tools has been exciting healthcare for decades. These tools have the potential to provide valuable insights on diagnosis, treatment options and prognosis. Unfortunately, many efforts failed in the migration process from research to clinical practice, largely due to the deficiency in the design of human-computer interaction and the consideration of collaborative nature of clinical workflow [122-124].

When narrowed down to particle therapy, one potential application of clinical decision system (CDS) is connecting past treatment decisions with current assessments, to help clinicians efficiently identify the optimum course of treatment, with regard to the selection of treatment modality and dose prescription. One exemplar study can be found in [125], in which Valdes et al. proposed an AI approach for early-stage lung and postoperative oropharyngeal cancer patients treated with photon or proton. A library of historical treatment plans and patient-specific feature sets were used to construct the classifiers. The authors claim that that based upon the learning curves, 45, 60, and 30 patients are needed for developing a sufficiently accurate classification model for early-stage lung, postoperative oropharyngeal (photon) and postoperative oropharyngeal (proton), respectively. Given both the complexity of the proposed task and the large degree of variations among patients, the abovementioned small data sizes seem to be overly optimistic.

### C. Replanning for adaptive therapy

This is a task interleaved with multiple tasks covered in this review, centering on how to make a treatment adaptive. The generation of an adapted plan involves two subtasks: fast dose calculation and fast plan optimization. The computational speed used to be a bottleneck, but GPUs have recently been deployed for speeding up [126-129]. The total time for generating a plan can be reduced down to 5 mins (MC dose calculation) [126] and 10 s (analytical dose calculation) [128]. How to optimize a plan can be devised along two avenues. One avenue is to moderately alter the position of Bragg peaks based on the latest geometry. For instance, it first adapts the energy of each pencil beam to the new water equivalent path length, and then re-optimizes beam weights using a standard optimization solver [127]. The other avenue is to produce a plan with newly added beamlets/spots (i.e. less constraints in re-planning), which can potentially improve dose conformity and spare OARs. This would be beneficial in cases where the relative distance between PTV and OAR significantly differs from the original plan [130, 131].

Range and dose verification in section IV is able to help form a closed loop in the adaptive workflow, a critical step to evince the dosimetric advantage of particle therapy over other modalities. Besides proton-induced secondary signals, this can also be achieved utilizing log-files [128, 132, 133], similar to the studies in section VI.

In our view, being "adaptive" comprise both before- and after-delivery adaptation. To be specific, if the anatomy has not changed above a threshold, the delivery can proceed as planned and no adaptation is needed. If the anatomy does change noticeably, the steps of dose influence recalculation and plan re-optimization steps are required before delivery. Finally, after-delivery dose verification can be performed to check whether fine-tuning spot weights between fractions is necessary. When synthesizing the abovementioned issues altogether, two intriguing questions arise. First, can an AI framework be used for online adaptation immediately after after-delivery verification, when the patient is still on the treatment couch? Second, can an AI framework be used for the fully automatic generation of adapted plans? More information about this topic can be found in a recently publish article [134]. As the field of particle therapy becomes more prevalent and sees an increasing number of AI-powered applications (e.g. dose calculation in section III, dose verification in section IV, CBCT image guidance in section V), AI-powered replanning will be the next step.

## VII. COMMON CHALLENGES AND PITFALLS

In spite of the promising progresses summarized in this review, we have to point out that AI is not silver bullet and faces its own challenges. As models become so complicated, it gets more challenging to inspect how inputs/outputs have been manipulated, as well as to interpret and check results. A number of common pitfalls to be avoided can be found in a lay article [135]. Below we briefly discuss three challenges and pitfalls highly prevalent to the field of radiotherapy (both particle and photon therapy), largely based on our own experiences.

1) *Generate and split data appropriately.* Compared to other fields such as computer vision and natural language processing, data scarcity is a huge bottleneck in particle therapy as encountered in a majority of examples we discussed so far. The limited size of dataset may cause several well-known problems, such as overfitting. To address such a limitation, data augmentation through MC simulation has often been used. How to effectively generate sufficient datsets is critical, and the answer will be both task-specific and model-specific. Three commitment issues related to data generation should also be considered. First, whether MC simulation produces the same (or comparable) results as would be expected in practical settings (e.g. system set-up, beam profile, image acquisition protocols), should be carefully examined. Second, as its name suggests, a patient-specific task (e.g. dose calculation, dose verification, QA) demands data generation per patient basis. How to balance a trade-off between being patient-specific and obtaining good generalization capability is important, since the latter one will have larger sample complexity and higher efficiency of data utilization. Transfer learning can also be considered here. Third, AI practitioners typically split data into training, validation and test sets. The splitting may or may not be done in a completely random manner. Convergence behaviors and bias/variance tradeoff should be always reported. Careful consideration and trying different approaches are highly encouraged to make sure the research findings stand the test of time.

2) *Interpretation with respect to features and hidden variables.* This is an extremely important aspect in AI-related studies, but is often overlooked. From our own experiences, it greatly helps avoid the overfitting problem. A well-known story in the machine-learning field is the "tank problem" [135, 136]. Researchers developed a model to spot tanks in pictures provided by the military. The model found the tanks successfully in the test dataset but failed later with real pictures. Simply put, it was other features (e.g. morning light, clouds, etc.) that drove the model, instead of the presence of tanks. In practice, a good way to check this is to use the same model to predict other things. If it succeeds, the results and study design may be skeptical. Furthermore, by inspecting the hidden variables, a more in-depth comparison can be made among different models as exemplified in [69]. With regard to input features and their respective contributions, several rounds of cross-validation should be considered. For example, the AI model was deployed for dose verification with three inputs: activity profile, HU from CT images and stopping power [70]. The authors intentionally altered each input and evaluated its impact on the overall performance, a very useful way to better understand how the model works inside a "black box". In another study, the order of acoustic waveforms was randomly permuted to pinpoint whether the time difference and sequential information were actually utilized for identifying the location of Bragg peak [94, 95]. Many AI papers in the field of particle therapy fail to go one step further to perform such experiments.

3) *Starting with the simplest model.* It seems natural for researchers to go with complex models. But sometimes a basic model without the use of neural networks, or with just a shallow neural network, performs equally well as a deep neural network with many layers. The overall aim must be kept in mind. Otherwise, one will be either setting up an unnecessarily complicated model to solve a simple problem, or sometimes even the wrong problem. If the purpose is only to predict the Bragg peak, a basic forward neural network [68] or a regression classifier [80] will suffice. Unless the purpose is to perform 3D dose verification, there is no need to go with advanced models such as RNN or GAN [68-71, 80]. Trying a simple baseline model is also useful on several regards: better understanding the dataset, establishing a performance baseline, choosing correct figures-of-merit, and guiding ablation studies if necessary.

Researchers conducting AI studies in the field of particle therapy should familiarize themselves with these common challenges and pitfalls, and hold themselves to higher standards. The soundness checks and error measurements should always be conducted. A clear standard on how to perform and report research findings is desired, and so is the standard for reviewing and publish AI-related articles.

## VIII. CONCLUSION

Computational power, big data and advanced algorithms are coming together to unleash the power of AI in radiotherapy. Due to two distinct differences between photon therapy and particle therapy (beam interaction physics and beam delivery mode), different strategies should be devised accordingly. Based upon the examples surveyed in the review, the AI deployment in particle therapy will not be "old wine in new bottles", but has great potential to address a number of unique unmet needs for boosting both accuracy and efficiency. Data generation, result interpretation, incorporation of fundamental physical processes and rigorous validation, are four aspects as critical as the AI model development itself. Clear standards about how to conduct and publish AI results need to be clearly established in the community. A clear pathway lies ahead of us to push the limit of AI tools towards more effective particle therapy.